\documentclass{article}
\usepackage{times}
\usepackage{cite}
\usepackage{amssymb,amsmath}
\newcommand{\ii}{\mathrm{i}}

\newcommand{\dd}{\mathrm{d}}
\newcommand{\pd}{\partial}

\newcommand{\e}{\mathrm{e}}

\newcommand{\tr}{\mathop{\mathrm{tr}}\nolimits}

\newcommand{\Op}{\mathcal{O}}

\newcommand{\ft}[2]{{\textstyle\frac{#1}{#2}}}

\begin{document}

\title{Dilatation operator or Hamiltonian?}

\author{Corneliu Sochichiu\thanks{E-mail:~\textsf{sochichi@lnf.infn.it}}\\
Max-Planck-Instut f\"ur Physik (Werner-Heisenberg-Institut),\\ F\"ohringer Ring 6, D-80805 Munich, GERMANY\\
Institutul de Fizic\u{a} Aplicat\u{a},\\
str.Academiei, 5, MD-2028 Chi\c{s}in\u{a}u, MOLDOVA\\
Laboratori Nazionali di Frascati,\\ via Enrico Fermi 40, I-00044 Frascati (RM), ITALY
}

\maketitle                   
\begin{abstract}
  We analyze the situation when the Hamiltonian in field theory can be replaced by the dilatation operator.
\end{abstract}

\section{Introduction}

According to AdS/CFT correspondence \cite{Maldacena:1997re}, the dynamics of string theory on the AdS$_5\times S^5$ background corresponds to the dilatation flows of composite operators in $\mathcal{N}=4$ super Yang--Mills (SYM) theory on the four-dimensional Minkowski space which is the conformal boundary of $AdS_5$. (See \cite{Aharony:1999ti} for a review.)

In the context of AdS/CFT correspondence the scaling properties of SYM composite operators appear extremely important
and were extensively studied last years (see \cite{Beisert:2004ry} for a review).

Using the dilatation operator as the Hamiltonian, one can define a dynamical system \cite{Beisert:2003jj,Bellucci:2004fh,Sochichiu:2006uz,Sochichiu:2006yv}. The matrix description of anomalous dimensions appears to be most natural if the non-planar effects are taken into consideration.

The replacement of the Hamiltonian-based analysis of finite temperature gauge theory by the Dilatation operator based analysis is widely used since the pioneering work of Polyakov \cite{Polyakov:2001af}. Since that, it was used many times in the same context in \cite{Sundborg:1999ue,Aharony:2003sx,Aharony:2005bq,Harmark:2006di}, where it appears that the dilatation operator based analysis is much simpler than the analysis based on the original gauge theory Hamiltonian. In particular, in \cite{Sochichiu:2006uz} the analysis of the thermodynamical properties of the matrix models corresponding to dilatations was proposed, for which different phases distinguished by the scaling properties of thermodynamical potentials were identified. As it appears, in the stringlike confining phase one can solve the model introducing random walk variables \cite{Sochichiu:2006yv}.

In this note we investigate when a replacement of the Hamiltonian dynamics with the dilatation operator dynamics is possible without affecting the thermodynamical behavior.

In what concerns the four-dimensional $\mathcal{N}=4$ super Yang--Mills (SYM) theory which is conformal invariant this replacement is justified by the conformal invariance of the model: The conformal transformation mapping the $\mathbb{R}^4$ with the punched origin into the cylinder $R\times S^3$, transforms the dilatations into time evolution flows and, therefore, the dilatation operator takes the function of the  Hamiltonian.\footnote{Here we assume the Euclidean signature.} If it is the case the Hamiltonian and the Dilatation operator are related by a unitary transformation and have the same spectrum.

The interesting point is whether this equivalence can be extended beyond a conformal theory framework. For example a compact sector of $\mathcal{N}=4$ is not invariant with respect to the above transformation mapping between dilatation and time evolution. Hence, the conformal symmetry does not imply an equivalence for such a sector, but only for the whole theory.
A perspective would be to extend the analysis to general gauge theories.

In present paper we concentrate on simplest cases, namely on free field as well as on the SU(2) sector of $\mathcal{N}=4$ SYM.

The plan of the note is as follows. In the second sections we review the Hamiltonian analysis of the free compactified field theory while in the third section we consider the dynamical system arising from the dilatation operator of the free theory and compare two systems. In the fourth section we consider the SU(2) sector of $\mathcal{N}=4$ SYM and find respective dynamical sector in the Hamiltonian description, first in zero coupling limit and then switching the interaction on. At the end we give our conclusions.
\section{Hamiltonian dynamics}

Let us consider a gauge theory in the limit of zero coupling and compactify it down to $1+0$ dimensions. Depending on the nature and geometry of compactification the compactified action generically takes the following form:
\begin{equation}\label{sh}
  S_{H}=\int\dd t \tr\left\{ \ft12 (D_t\phi_a)^2-\ft12 m^2\phi_a^2\right\}+KK,
\end{equation}
where $\phi_a$, $a=1,\dots,M$ are the lowest Kaluza-Klein (KK) modes for all fields but the time component of the gauge field $A_0$ which enters through the covariant derivative:
\begin{equation}
  D_t\phi_a=\dot{\phi}_a+[A_0,\phi_a].
\end{equation}
The mass which can be either inherited from the higher dimensional theory either an artifact of compactification or result of both generally can be different for different fields, but we so far neglect this fact to remain in the framework of the simplest setup. The remaining terms denoted by $KK$ represent the contribution of the higher KK modes, which we will drop in the further analysis.

The action \eqref{sh} can be rewritten in terms of holomorphic field $\Phi_a$ and anti-holomorphic one $\bar{\Phi}_a$ the classical analog of raising an lowering operators,
\begin{equation}
  S_{H}=\int\dd t\tr\left\{
  \ii \bar{\Phi}_a D_t\Phi_a-m \bar{\Phi}_a\Phi_a
  \right\},
\end{equation}
where
\begin{equation}
  \Phi_a=\ft1{\sqrt{2m}}\left(\Pi_a-\ii m \phi_a\right),\qquad \bar{\Phi}_a=\ft1{\sqrt{2m}}\left(\Pi_a+\ii m \phi_a\right),
\end{equation}
and $\Pi_a$ is the canonical momentum conjugated to $\phi_a$: $\Pi_a=\dot{\phi}_a$.

Due to the simplicity of the model we can handle both Hamiltonian based and dilatation based approaches to confront them.

\section{Dilatation dynamics}

According to our ``simplism'' ideology, we assign to all original fields classical dimension one. Due to the absence of interaction the dimension of composite operators is the sum of constituent fields. The allowed operators are gauge invariant polynomials of the fields and their derivatives. To simplify further we restrict ourself to ultra-local operators only i.e. no derivative insertions.\footnote{Our assumption is that the ultra-local sector is related to the lowest KK modes while inclusion of derivatives puts in the game the higher ones.} Thus, a typical operator looks like,
\begin{equation}\label{comp}
  \Op(\{a\},\{b\},\dots,\{c\})=
  \tr \Phi_{a_1}\dots \Phi_{a_{L_1}}\tr \Phi_{b_1}\dots \Phi_{b_{L_2}}\dots\tr\Phi_{c_1}\dots \Phi_{c_{L_k}},
\end{equation}
where $L=L_1+L_2+\dots+L_k$ is the the dimension of the operator $\Op(\{a\},\{b\},\dots,\{c\})$ and, in order not to abuse the notations we use the capital $\Phi$ for the higher dimensional field. The above system can be mapped to a gauged matrix oscillator where the operators of the above type will correspond to the gauge invariant states while the dilatation operator,\footnote{Note a slightly distorted notations with respect to the standard oscillator ones: the ordinary symbols are representing raising operators, while the checked symbols are lowering ones.}
\begin{equation}
  \Delta=\tr\Phi_a\check{\Phi}_a,\qquad (\check{\Phi}_a)_{m}{}^{n}=\frac{\pd}{\pd (\Phi_a)_{n}{}^{m}},
\end{equation}
becomes Hamiltonian.

The action corresponding to this system is \cite{Bellucci:2004fh,Sochichiu:2006uz},
\begin{equation}\label{sd}
  S_{\Delta}=\int\dd t \tr\left\{
  \ii \bar{\Phi}_a D_t\Phi-\bar{\Phi}_a\Phi_a
  \right\},
\end{equation}
where the covariant derivative is $D_t\Phi_a=\dot{\Phi}_a+[A_0,\Phi_a]$. Now the matrix $\Phi_a$ is a complexification of the original matrix, i.e. the matrix fields where promoted from Hermitian to non-Hermitian matrices. Let us recall that the meaning of the time parameter $t$ in this action is the scale factor rather the true time.

Comparing the dilatation action \eqref{sd} to the action of the compactified model in the form \eqref{sh}, we see that the models coincide up to a scale factor given by the mass parameter $m$. If the non-zero mass term is one induced by the compactification, one is able to adjust the compactification radius in such a way to fit the value $m=1$. In this case the conclusion is that both the Hamiltonian based and the Dilatation operator based dynamics are essentially equivalent.

The matrix oscillator described by either \eqref{sh} or \eqref{sd} was thoroughly studied, for details we can refer the reader to the literature: \cite{Sundborg:1999ue,Aharony:2003sx,Aharony:2005bq,%
Sochichiu:2006uz,Sochichiu:2006yv}.

\section{SU(2) sector of $\mathcal{N}=4$ SYM}

In the case of SU(2) sector of $\mathcal{N}=4$ SYM one should restrict to holomorphic operators \eqref{comp} consisting of two complex fields $\Phi_a$, $a=1,2$, but not of their complex conjugate. The dilatation analysis does not change much, except that in this case no complexification of $\Phi_a$ occurs: the composite operators already span the Hilbert space of gauged matrix oscillator (see the Appendix of \cite{Sochichiu:2006uz}). As a result one ends up with \emph{half} of the degrees of freedom comparing to the previous case. The role of the anti-holomorphic field is merely the canonical conjugate to the holomorphic one.

Let us turn to the respective compactified model in the limit $g_{\rm YM}\to 0$. The minimal part of the SYM action in zero coupling limit which contains the complex fields $\phi_a$ looks like follows
\begin{equation}\label{su2}
  S_{H}=\int\dd t \tr\left\{ D_t\bar{\phi}_a D_t\phi_a-m^2\bar{\phi}_a\phi_a
  \right\}.
\end{equation}

As one can see, the compactified model is ``too large'': a pair of fields $(\bar{\phi}_a,\phi_a)$ describe \emph{two} oscillators instead of just one. This corresponds to composite operators containing insertions of both $\phi_a$ and $\bar{\phi}_a$, i.e. the action \eqref{su2} the SO(4)=SU(2)$\times$SU(2) sector rather just SU(2) one. Moreover, because of the reality condition for the action it seems impossible to separate $\phi_a$ from its complex conjugate. The action \eqref{su2}, however, reduces to \eqref{sd} in some special limit.

Let us consider a synchronous rotation of $\phi_a$-planes:
\begin{equation}\label{rot}
  \phi_a\to \Phi_a=\sqrt{\omega}\e^{\ii \omega t}\phi_a,\qquad
  \bar{\phi}_a\to\bar{\Phi}_a=\sqrt{\omega}\e^{-\ii\omega t}\bar{\phi}_a,
\end{equation}
and consider a large angular velocity $\omega$. In this limit the action \eqref{su2} takes the form,
\begin{equation}
  S_{\omega\to\infty}=
  \int\dd t\tr\left\{
  \ii \bar{\Phi}_a D_t \Phi_a-m_{\omega}\bar{\Phi}_a\Phi_a,
  \right\}+O (1/\omega),
\end{equation}
where $m_{\omega}=\omega+m^2/\omega$. As a conclusion the dilatation operator restricted to holomorphic composite operators represents the spectrum of rotating field in units of angular velocity. Moreover, it appears that the presence of mass term is not very important in this case as a mass is generated by the rotation and we can put $m=0$. Let us note that this trick is similar to one used to identify strings dual to SU(2) spin chain solutions in \cite{Kruczenski:2003gt}. In fact, it is convenient to express the action in units of $\omega$ which means that we are considering processes with energies much less than one.

It is interesting to note that the same purpose can be reached by imposing a set of second class constraints. To do this let us firs rewrite the action in the first order formalism,
\begin{equation}\label{mlessh}
  S_{H}=\int\dd t\tr \left\{
  \bar{\pi}_a D_t\phi_a+\pi_a D_t\bar{\phi}_a-\bar{\pi}_a\pi_a
  \right\}.
\end{equation}
After imposing following constraints:
\begin{equation}\label{cons}
  \varphi=\pi_a+\ii \phi_a\approx 0,\qquad  \bar{\varphi}=\bar{\pi}_a-\ii \bar{\phi}_a\approx 0,
\end{equation}
the action \eqref{mlessh} becomes equivalent to \eqref{sd}.
The desired form is obtained by rewriting the action \eqref{mlessh} in terms of the following on-shell coordinates:
\begin{equation}
  \Phi_a\equiv\phi_a\approx \ii \pi_a , \qquad \bar{\Phi}_a\equiv\bar{\phi}_a\approx -\ii \bar{\pi}_a.
\end{equation}

\section{Switching the interaction on}

In this simple example of SU(2) sector of $\mathcal{N}=4$ SYM theory we know enough data to see what happens when the interaction is switched on. Switching on the coupling $g_{\rm YM}$ results in the addition of the following interaction term to the action \eqref{su2},
\begin{equation}\label{shg}
  S_{H}=\int\dd t \tr\left\{ D_t\bar{\phi}_a D_t\phi_a-m^2\bar{\phi}_a\phi_a
  -\ft{g_{\rm YM}^2}{2}\left(
  [\bar{\phi}_a,\bar{\phi}_b][\phi_a,\phi_b]+
  [\bar{\phi}_a,\phi_b][\phi_a,\bar{\phi}_b]
  \right)
  \right\}.
\end{equation}

It may seem that the application of any of described above reduction procedures does not have any effect on the interaction term since on one hand it is invariant with respect to rotations \eqref{rot} and on the other is already on-shell with respect to the constraints \eqref{cons}, but this does not take into account the metamorphosis of the gauge symmetry. In either of the above reduction procedures the Gauss law constraint changes to the following form,
\begin{equation}
  G=[\bar{\phi}_a,\phi_a]\approx 0.
\end{equation}
The new Gauss law constraint allows one to rewrite the last term of \eqref{shg} in the form as follows,
\begin{equation}
  \tr[\bar{\phi}_a,\phi_b][\phi_a,\bar{\phi}_b]\approx
  \tr[\bar{\phi}_a,\bar{\phi}_b][\phi_a,\phi_b],
\end{equation}
which is identic to the remaining commutator term in \eqref{shg}. This allows one to write the reduced action action in the following form,
\begin{equation}
  S_{H}=\int\dd t\tr\left\{
  \ii \bar{\phi}_a D_t \phi_a-\bar{\phi}_a\phi_a
  -g_{\rm YM}^2[\bar{\phi}_a,\bar{\phi}_b][\phi_a,\phi_b]
  \right\}.
\end{equation}
Up to  a redefinition of coupling $g_{\rm YM}^2\to g_{\rm YM }^2/16\pi^2$ this is completely equivalent to the matrix model describing the dilatation dynamics considered in \cite{Beisert:2002ff,Bellucci:2004fh,Sochichiu:2006uz,Sochichiu:2006yv}. Again this redefinition of the interaction coupling can be interpreted as an adjusting of scales.
\section{Discussion and Outlook}

In this paper we tried to answer the question: when one can replace the dynamics of a quantum field theory by one generated by the action of the dilatation operator. So far we considered the simplest cases: one of free field and one of SU(2) sector of $\mathcal{N}=4$ SYM theory in one-loop approximation.

As about free theory we have shown that in the case of a  massive field a complete equivalence of the spectra and therefore of thermodynamical quantities can be achieved by adjusting the scales, which makes possible replacement of Hamiltonian dynamics by the dilatational.
As soon the relation concerns quantum oscillators the conclusion does not come too surprising given the rich symmetry of the system. In particular, in holomorphic coordinates which are classical analogues of rising and lowering operators the evolution in imaginary time is equivalent to scaling transformation which is induced by the dilatation operator.

In the case of SU(2) sector of SYM theory, the respective sector in the Hamiltonian description can be identified by use of either fast rotating limit or imposing particular second class constraints on the Hamiltonian corresponding to SO(4) sector. Again, the correspondence is attained by adjusting the scales. At this stage the physical meaning of the construction is not very clear. Also it is not clear how to systematically extend the analysis to higher loops or even how to include the two-loop contribution.

On the other hand, one can relax the conditions asking for the thermodynamical equivalence in the large $N$ limit only. This may extend considerably the class of models for which the substitution of the Hamiltonian statistical mechanics with the dilatation one can be harmlessly applied. In any case this issue deserves a further study.


\subsubsection{Acknowledgement}
  This work was started in Max-Planck-Institut in Munich. I am grateful to Dieter Luest for warm hospitality and support of my research. I also benefited  from useful discussions with Jeong-Hyuck Park.

  This work was supported by RTN project ``ForcesUniverse" MRTN-CT-2004-005104

%
%
\end{document}